\documentclass[%
reprint,
superscriptaddress,
amsmath,amssymb,
aps,
prl,
]{revtex4-1}

\usepackage{tcolorbox}
\usepackage{graphicx}
\usepackage{dcolumn}
\usepackage{bm}
\usepackage{epstopdf}

\begin{document}

\preprint{APS/123-QED}

\title{Eckhaus Instability in the Fourier-Domain Mode Locked Fiber Laser Cavity}

\author{Feng Li}
\affiliation{Photonic Research Centre, Department of Electronic and Information Engineering, the Hong Kong Polytechnic University, Hung Hom, Hong Kong SAR, China}%
\affiliation{The Hong Kong Polytechnic University Shenzhen Research Institute, Shenzhen, 518057, China}

\author{K. Nakkeeran}%
\affiliation{School of Engineering, Fraser Noble Building, University of Aberdeen, Aberdeen AB24 3UE, UK}

\author{J. Nathan Kutz}
\affiliation{Department of Applied Mathematics, University of Washington, Seattle, WA 98195-2420, USA}

\author{Jinhui Yuan}
\email{yuanjinhui81@163.com}
\affiliation{Photonic Research Centre, Department of Electronic and Information Engineering, the Hong Kong Polytechnic University, Hung Hom, Hong Kong SAR, China}
\affiliation{State Key Laboratory of Information Photonics and Optical Communications, Beijing University of Posts and Telecommunications, P.O. Box 72, 100876 Beijing, China}

\author{Zhe Kang}
\affiliation{Photonic Research Centre, Department of Electronic and Information Engineering, the Hong Kong Polytechnic University, Hung Hom, Hong Kong SAR, China}

\author{Xianting Zhang}
\affiliation{Photonic Research Centre, Department of Electronic and Information Engineering, the Hong Kong Polytechnic University, Hung Hom, Hong Kong SAR, China}

\author{P. K. A. Wai}
\email{alex.wai@polyu.edu.hk}
\affiliation{Photonic Research Centre, Department of Electronic and Information Engineering, the Hong Kong Polytechnic University, Hung Hom, Hong Kong SAR, China}
\affiliation{The Hong Kong Polytechnic University Shenzhen Research Institute, Shenzhen, 518057, China}

\date{\today}

\begin{abstract}
High frequency fluctuation in the optical signal generated in Fourier-Domain Mode Locked fiber laser (FDML-FL), which is the major problem and degrades the laser performance, is not yet fully analyzed or studied. The basic theory which is causing this high frequency fluctuation is required to clearly understand its dynamics and to control it for various applications. In this letter, by analyzing the signal and system dynamics of FDML-FL, we theoretically demonstrate that the high frequency fluctuation is induced by the intrinsic instability of frequency offset of the signal in cavity with nonlinear gain and spectral filter. Unlike the instabilities observed in other laser cavities this instability is very unique to FDML-FL as the central frequency of the optical signal continuously shifts away from the center frequency of the filter due to the effects like dispersion and/or nonlinearity. This instability is none other than the Eckhaus instability reported and well studied in fluid dynamics governed by real Ginzburg-Landau equation.
\end{abstract}

\maketitle
\setlength{\textfloatsep}{10pt plus 3pt minus 2pt}
\setlength{\abovecaptionskip}{5pt} 
\setlength{\belowcaptionskip}{2pt} 

Nonlinear systems either naturally existing or manmade exhibit fascinating dynamics and have various applications in different fields. Development of appropriate and complete theoretical models of these nonlinear systems play vital role in understanding their dynamics and to control them for various applications. Nonlinear systems are governed either by nonlinear ordinary differential equations or nonlinear partial differential equations (NPDEs). Some NPDE governs more than one kind of systems. Complex Ginzburg Landau equation (CGLE) is one such very famous NPDE which governs nonlinear dynamical systems from different fields like fluids, optics, superconductivity and Bose-Einstein Condensate. The derivatives of CGLE like real Ginzburg Landau equation (RGLE) and the family of nonlinear Schr\"odinger equation (NLSE) also widely appear as system model equations across different fields. As a system dynamical equation governs more than one type of systems it became very common to utilize the results, studies and analyzes obtained in one field to the corresponding other fields systems governed by the same model equation. This helps in understanding the dynamics of the systems governed by the same dynamical equation at a faster pace and also to make further appropriate modifications and/or to achieve complete theoretical model.

In this work we consider the Fourier-Domain Mode Locked fiber laser (FDML-FL) cavity which is a long cavity wavelength swept laser source and has a very important application in optical coherent tomography (OCT) \cite{Huang1991,Fujimoto2003,Robles2011}. FDML-FL was experimentally demonstrated for the first time in 2005 \cite{Huber2005,Huber2006}. To avoid the rebuilding of laser signal from the spontaneous emission, which intrinsically limits the sweeping speed of wavelength swept lasers, a long fiber delay line was introduced into the cavity of FDML-FL to buffer the entire wavelength sweeping signal. With this kind of FDML-FL cavity, the wavelength sweeping speed can be enhanced by one to two orders to MHz level \cite{Klein2012,Wieser2010}. Although FDML-FL has been successfully deployed in the OCT systems as the swept source, the performance of the FDML-FL is limited by its large instantaneous linewidth \cite{Biedermann2009,Adler2011,Wieser2012}, which is in fact the high frequency fluctuations of the signal waveform. In order to improve the performance and to understand the dynamics of the FDML-FL, in this letter using an appropriate theoretical model we report the complete working mechanism and the intrinsic reason of such fluctuations in the FDML-FL cavity.

In an FDML-FL, the fiber dispersion degrades the quality of the signal due to the mismatch between the filter sweeping period and the different round trip time of different parts of the sweeping signal in the cavity \cite{Biedermann2009,Jirauschek2009}. The linewidth of the signal is further increased by the nonlinearity of the fiber and the linewidth enhancement factor of the semiconductor optical amplifier (SOA) \cite{Todor2012}. To investigate the dynamics of the signal in the FDML-FL cavity, a theoretical model in a co-moving frame of the filter has been proposed as \cite{Jirauschek2009,Todor2012}
\begin{eqnarray}
\label{eq:1_FDMLfull}
{\partial_z}u&&=g(u,\omega_s)(1-i\alpha)u-\sigma (\omega _s)u-a( i{\partial_t} )u\nonumber \\
&&+i{{D}_{2}}\omega _{s}^{2}(t)u+i{{D}_{3}}\omega _{s}^{3}(t)u+i\gamma {{\left| u \right|}^{2}}u-i{{D}_{2}}\partial _{t}^{2}u,
\end{eqnarray}

\noindent where $u$ is the amplitude in filter frame defined as $u=A\exp(i\int^t{\omega_s(t')dt'})$, $\omega_s$ is the instantaneous center frequency of the sweeping filter and $A$ is the complex amplitude of the signal in lab frame. This model includes the effects such as dispersion, nonlinearity, linewidth enhancement factor and frequency filtering. Equation (\ref{eq:1_FDMLfull}) is widely used in the numerical simulation of FDML-FL. Through numerical simulations of the FDML-FL, there are always high frequency fluctuations in the signal waveform\cite{Jirauschek2009,Todor2012}. These fluctuations appear predominantly on the desired signal and hence usually the waveform is artificially smoothened over a long period of time scale \cite{Jirauschek2009,Todor2012}. To understand the intrinsic reason for this like noise fluctuation, we use a simplified model. Utilizing the Wentzel--Kramers--Brillouin (WKB) analysis \cite{Kevorkian1996}, we find that the nonlinear phase shift and in-band dispersion are playing minor role and those terms can be neglected in the first order approximation \cite{Li2013}. The large linewidth enhancement factor, which is a special effect of SOAs, can also be neglected in a general laser cavity. Also neglecting the wavelength dependence of the gain and considering a typical Gaussian spectral filter with bandwidth $B$, Eq.~(\ref{eq:1_FDMLfull}) becomes
\begin{eqnarray}
\label{eq:3_WKBGaussian}
\!\!\!\!\!\!\!{{\partial }_{z}}u=&&gu-\sigma u+\tfrac{1}{2}{{B}^{-2}}\partial _{t}^{2}u+i{D_2}\omega _s^2(t)u+i{D_3}\omega _s^3(t)u,
\end{eqnarray}

\noindent where $g(u)=g_0/(1+|u|^2/I_{sat})$ is the saturated gain. It should be noted that such fast response gain saturation model is valid not only in the case with a fast recovery gain, but also in the cavity with a slow recovery gain and a fast response nonlinear loss element such as nonlinear optical loop mirror, nonlinear polarization rotation device. The simulation results of the simplified model of Eq.~(\ref{eq:3_WKBGaussian}) capture most of the signal dynamics of the FDML-FL system, especially the high frequency intensity fluctuation of the waveform. Now, the gain saturation factor is expressed in Taylor series of $|u|^2$ and keeping only the first order term, Eq.~(\ref{eq:3_WKBGaussian}) reduces to
\begin{eqnarray}
\label{eq:4_nonlineargain}
{\partial_z}u=&&(g_0-\sigma)u-{g_0}I_{sat}^{-1}{|u|^2}u+\tfrac{1}{2}{B^{-2}}\partial_t^2u\nonumber \\ 
 &&+i{D_2}\omega _s^2(t)u+i{D_3}\omega _s^3(t)u,
\end{eqnarray}

\noindent which can be normalized to
\begin{eqnarray}
\label{eq:5_RGLEC}
{{\partial }_{Z}}U=U-{|U|^2}U+\partial _T^2 U+i{{\epsilon }^{-1}}C(\epsilon T)U,
\end{eqnarray}

\noindent where $C(\epsilon T)=\epsilon \left[ {{S}_{2}}\Omega _{s}^{2}(T)+{{S}_{3}}\Omega _{s}^{3}(T) \right]$ is the combined phase term caused by the dispersion and ${S}_{2}, {S}_{3}$ are the normalized dispersion coefficients. The time scaling factor $\epsilon$ is defined as the inverse of the round trip time of the laser cavity. 

Equation~(\ref{eq:5_RGLEC}) is a RGLE with a chirp phase term $C$ contributed by the dispersion in the FDML-FL cavity. In an ideal cavity without dispersion, Eq.~(\ref{eq:5_RGLEC}) will reduce to a standard RGLE as
\begin{eqnarray}
\label{eq:7_RGLE}
{{\partial }_{Z}}U=U-{{\left| U \right|}^{2}}U+\partial _{T}^{2}U.
\end{eqnarray}

\noindent RGLE has been extensively studied in fluid dynamics. Most importantly, a set of stationary solutions are available for the system governed by the RGLE which are single frequency continuous waves. The stationary solution with normalized angular frequency $\Omega$ is
\begin{eqnarray}
\label{eq:8_solutionRGLE}
U=\sqrt{1-{\Omega ^2}}{e^{-i\Omega T}},
\end{eqnarray}

\noindent which are nontrivial in the frequency region $|\Omega|<1$. But the stationary solutions are unstable when $\Omega^2>1/3$, which is known as Eckhaus instability. Eckhaus instability was first discussed in the modeling of convection in fluidic systems governed by RGLE in 1960s \cite{Eckhaus1965}. After a very short time, the interest on the instability has moved to the system described by the CGLE, which is the modulation instability, where the Eckhaus instability can be treated as a reduced case with zero imaginary terms \cite{Zakharov2009}. As a reduced form of CGLE, NLSE has attracted more attention than the RGLE because of the existence of the analytical solitary solution, especially after soliton was reported in optical fibers by Hasegawa and Tappert \cite{Hasegawa1973}. With NLSE, the modulation instability of optical continuous wave has been very well studied. In a system with dissipative and gain elements, such as a laser cavity, the more general system equation, CGLE is adapted to model the nonlinear pulse dynamics in the cavity \cite{Akhmediev1997,Akhmediev2001,Komarov2005}. But in all those studies, the nonlinear phase shift and dispersion were considered as the dominant effects in pulse shaping and stable propagation. Although RGLE is seldom used to describe an optical system, Eckhaus instability has also been discussed when considering the spatial effects in lasers \cite{Elgin1986,Plumecoq2001,Wolfrum2006}. In an one dimensional cavity, where spatial effects are not considered, the Eckhaus instability never appeared as a dominating effect.

In the laser cavities described by Eq.~(\ref{eq:5_RGLEC}), there are no stationary solutions because of the existence of a single nonzero phase term. Only when an initial field is given, the evolution of the signal can be solved for Eq.~ (\ref{eq:5_RGLEC}). To solve Eq.~(\ref{eq:5_RGLEC}), we introduce an amplitude-phase form $\bar{U}=U\exp [-i{{\epsilon }^{-1}}C(\epsilon T)Z]$ and split the temporal dynamics by fast time $t_1=T$ and slow time $t_2=\epsilon T$, then the zero-th order governing equation of $\bar{U}$ can be obtained as
\begin{eqnarray}
\label{eq:11_diffUbarorder0}
{{\partial }_{Z}}\bar{U}=&&\bar{U}-{{\left| {\bar{U}} \right|}^{2}}\bar{U}+\partial _{{{t}_{1}}}^{2}\bar{U}-{{{C}'}^{2}}{{Z}^{2}}\bar{U}+2i{C}'Z{{\partial }_{{{t}_{1}}}}\bar{U},
\end{eqnarray}

\noindent where ${C}'={{\partial }_{\epsilon T}}C$. If a slowing varying signal is considered, the evolution of the signal along $Z$ governed by Eq.~(\ref{eq:11_diffUbarorder0}) can be written as
\begin{eqnarray}
\label{eq:12_solutionUbarorder0}
{{\bar{U}}(Z)}=\frac{\exp ( -i{{\bar{\Omega }}_0}(t_2){t_1} )}{\sqrt{I_0^{-1}{e^{Q(0)-Q(Z)}}+2\int_0^Z{{e^{Q(x)-Q(Z)}}dx}}},
\end{eqnarray}

\noindent where $I_0$ and $\bar{\Omega}_0(t_2)$ are the intensity and instantaneous frequency of  $\bar{U}$ at $Z=0$, and $Q(Z)=2\int_0^Z{[ 1-{({\bar{\Omega}_0}(t_2)-C'x)}^2 ]dx}$. The stability of the solution (\ref{eq:12_solutionUbarorder0}) can be investigated through linear stability analysis. A perturbation $a(t_1,t_2)$ is applied to the solution at $Z = Z_0$ as $W({{Z}_{0}})=\bar{U}({{Z}_{0}})(1+a)$. The evolution of the perturbed solution is assumed as
\begin{eqnarray}
\label{eq:14_defPerturbation}
W(Z,{{t}_{1}},{{t}_{2}})=\bar{U}(Z,{{t}_{1}},{{t}_{2}})[1+e^{\Lambda (Z,{{t}_{2}})}a({{t}_{1}},{{t}_{2}})],
\end{eqnarray}

\noindent where ${{\partial }_{Z}}\Lambda =\lambda (Z,{{t}_{2}})$ indicates the growing speed of the perturbation. If $\lambda (Z>{{Z}_{0}},{{t}_{2}})>0$, the perturbation at time point $t_2$ gets continuously amplified and the solution becomes unstable. By substituting Eq.~(\ref{eq:14_defPerturbation}) into Eq.~(\ref{eq:11_diffUbarorder0}), and using the solution described by Eq.~(\ref{eq:12_solutionUbarorder0}), where ${{\partial }_{{{t}_{1}}}}{{\bar{U}}}=-i{{\bar{\Omega }}_{0}}({{t}_{2}}){{\bar{U}}}$, the governing equation of  the perturbation is written as
\begin{eqnarray}
\label{eq:16_govpertsim}
\lambda a=&&-I\left( a+{{a}^{*}} \right)+\partial _{{{t}_{1}}}^{2}a+2i {\Omega }_{i}{{\partial }_{{{t}_{1}}}}a,
\end{eqnarray}

\noindent where $I$ is the intensity and ${{\Omega }_{i}}={C}'Z-{{{\bar{\Omega }}}_{0}}({{t}_{2}})$ is the instantaneous frequency of $U$. Considering the coupling between the conjugated fields, the field is written as
\begin{eqnarray}
\label{eq:17_ansatzpert}
a({{t}_{1}},{{t}_{2}})={{\alpha }_{k}}({{t}_{2}})\exp (-ik{{t}_{1}})+{{\beta }_{k}}({{t}_{2}})\exp (ik{{t}_{1}}),
\end{eqnarray}

\noindent where $k>0$ is the mode number of the perturbation. Substituting Eq.~(\ref{eq:17_ansatzpert}) into Eq.~(\ref{eq:16_govpertsim}), it is easy to find that  $\alpha_k$ and $\beta_k$ have nonzero solutions only when
\begin{eqnarray}
\label{eq:19_deteigen}
\left| \begin{matrix}
   I+{{k}^{2}}+2k{{\Omega }_{i}}+\lambda  & I+\lambda   \\
   I+\lambda  & I+{{k}^{2}}-2k{{\Omega }_{i}}+\lambda   \\
\end{matrix} \right|=0,
\end{eqnarray}

\noindent which has solutions $k=0$, or $\lambda =2\Omega _{i}^{2}-I-0.5{{k}^{2}}$ for $k>0$. Clearly, $\lambda$ will have negative values for all $k>0$ modes only when $2\Omega _{i}^{2}-I<0$, which is the criterion of the stability of the solution. With the solution described by Eq.~(\ref{eq:12_solutionUbarorder0}), the stability condition is
\begin{eqnarray}
\label{eq:22_eigenmax}
&&{{\lambda }_{max}}=2{{\left( {C}'Z-{{{\bar{\Omega }}}_{0}} \right)}^{2}}\nonumber\\
&&-{{\left[ I_{0}^{-1}{{e}^{Q(0)-Q(Z)}}+2\int_{0}^{Z}{{{e}^{Q(x)-Q(Z)}}dx} \right]}^{-1}}<0.
\end{eqnarray}

As the simplest case, where $C'=0$, solution (\ref{eq:12_solutionUbarorder0})  reduces to a quasi-stationary solution
\begin{eqnarray}
\label{eq:23_chirpUbar}
{{\bar{U}}_{0}}=\sqrt{1-\bar{\Omega }_{0}^{2}({{t}_{2}})}\exp \left[ -i{{{\bar{\Omega }}}_{0}}({{t}_{2}}){{t}_{1}} \right].
\end{eqnarray}

\noindent Also, $Q$ is simplified to $Q(Z)=2Z\left[ 1-\bar{\Omega }_{0}^{2}({{t}_{2}}) \right]$, and stability criterion of the eigenvalue changes to ${{\lambda }_{max}}=3\bar{\Omega }_{0}^{2}({{t}_{2}})-1<0$, which is exactly the condition for Eckhaus instability. Especially when ${{\bar{\Omega }}_{0}}({{t}_{2}})$ is a constant, the solution becomes the stationary solution of Eq.~(\ref{eq:8_solutionRGLE}). Such stationary solution can be found in laser cavities modeled by RGLE. A nonzero relative frequency offset ${{\bar{\Omega }}_{0}}$ can be introduced either by a frequency shifter, or a fast tuning of the spectral filter in the cavity. The stability of the stationary signal depends on the offset frequency ${{\bar{\Omega }}_{0}}$, which can be divided into three regions, $\bar{\Omega }_{0}^{2}<{1}/{3}$, ${1}/{3}\le \bar{\Omega }_{0}^{2}\le 1$ and $\bar{\Omega }_{0}^{2}>1$. The frequencies $\bar{\Omega }_{0}^{2}={1}/{3}\;$ and $\bar{\Omega }_{0}^{2}=1$ stand for the critical point to trigger the Eckhaus instability and the threshold frequency of positive net gain, respectively.

\begin{figure}[h]
\includegraphics[width=3.4in]{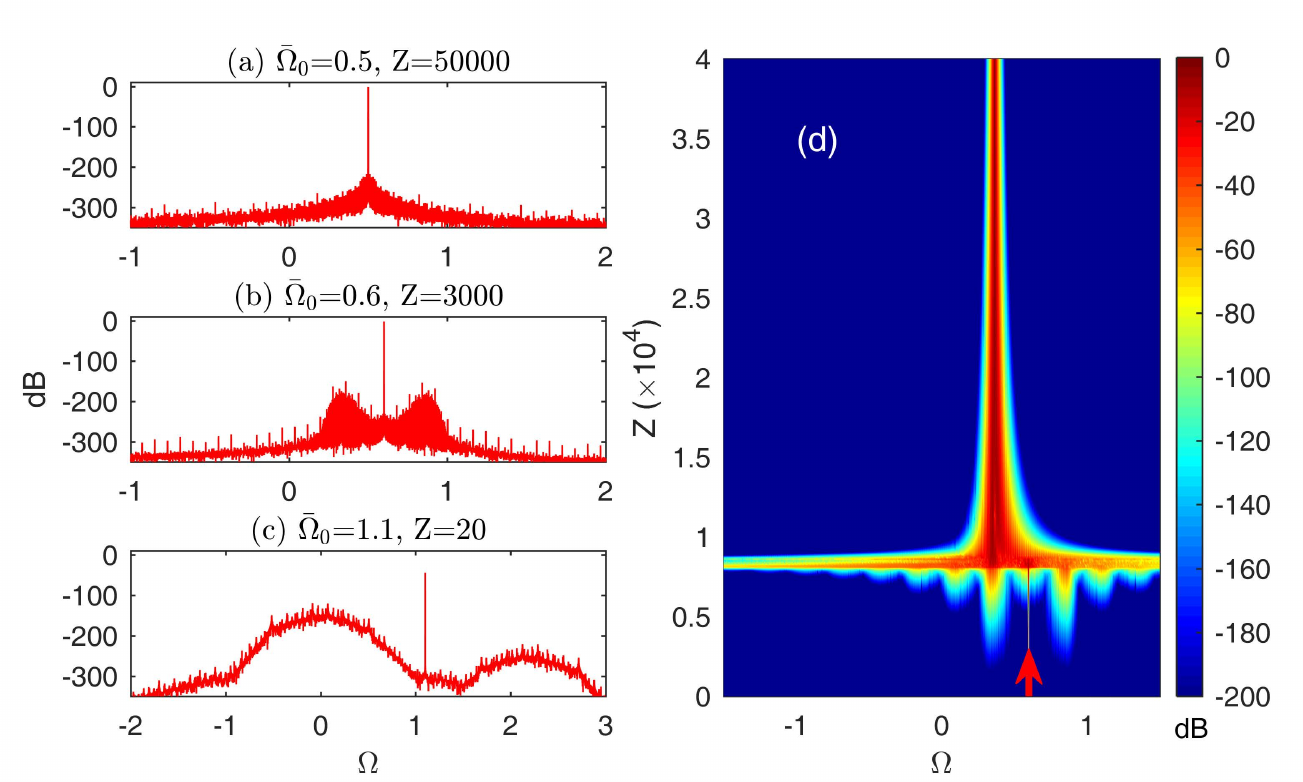}
\caption{\label{fig:1_sidebands} The spectra of the stationary signals after circulating in the cavity with frequencies in (a) stable region, (b) Eckhaus instability region and (c) net loss region. (d) The spectrum dynamics of a stationary signal with ${{\bar{\Omega }}_{0}}=0.6$, where Eckhaus instability is triggered. The red arrow indicates the frequency of the stationary signal.}
\end{figure}

Figure \ref{fig:1_sidebands} shows the spectra of signals with frequencies in the three different regions. In Fig.~\ref{fig:1_sidebands}(a), the frequency ${{\bar{\Omega }}_{0}}=0.5$, which is in the stable region of $\bar{\Omega}_0^2<1/3$, the amplitude of the signal is $-0.866$ dB. The solution is stable and the sidebands have not grown at $Z=50000$. When ${{\bar{\Omega }}_{0}}$ increases to $0.6$, which has passed the critical point $\bar{\Omega }_{0}^{2}={1}/{3}$, as shown in Fig.~\ref{fig:1_sidebands}(b), sidebands are generated on both sides of the signal at $Z=3000$. In this region, although the signal is unstable, the original signal decays very slowly until the sidebands grow to relatively large value. To explain the mechanism of the growth of sidebands and frequency switching, we show the spectral evolution of the signal in Fig.~\ref{fig:1_sidebands}(d). Note that the signal switching occurs in a very short distance where higher order sidebands are also excited. Eventually, the sideband at the low frequency side, which experiences higher net gain, becomes the dominant mode and replaces the original single frequency signal. The new signal is stable as it is in the region of $\bar{\Omega }_{0}^{2}<{1}/{3}$. When ${{\bar{\Omega }}_{0}}$ is further increased to $\bar{\Omega }_{0}^{2}>1$, as shown in Fig.~\ref{fig:1_sidebands}(c), the solution defined in Eq.~(\ref{eq:23_chirpUbar}) does not exist anymore and the single frequency signal gets quickly attenuated. At the same time, new signal builds up from noise.

\begin{figure}[t]
\includegraphics[width=3.4in]{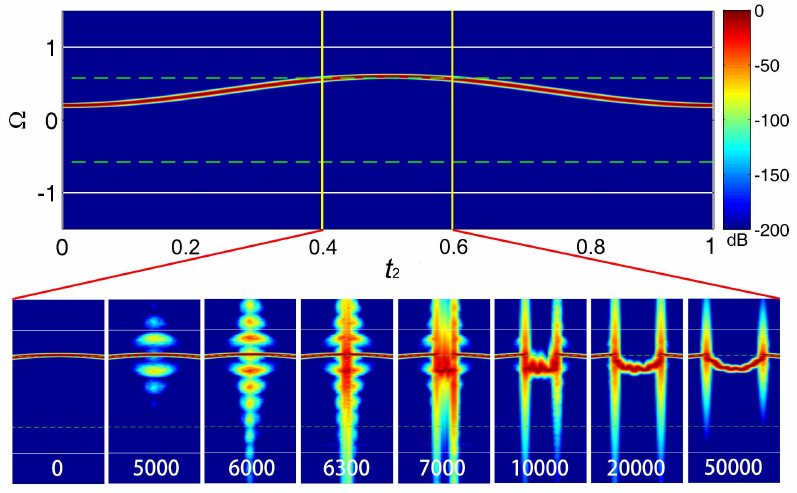}
\caption{\label{fig:3_chirp_evolution} The spectrograms at different $Z$ of the signal with frequency ${{\bar{\Omega }}_{0}}({{t}_{2}})=0.4-0.2 \cos (2\pi{t}_{2})$. The top figure shows the full spectrogram of signal at $Z=0$. The bottom figures show the evolution of the spectrogram in the region of $0.4<t_2<0.6$ from $Z=0$ to 50000. The dashed lines in the maps indicate the boundaries to trigger Eckhaus instability.}
\end{figure}

In the FDML-FL cavity, the dispersion will introduce a time varying chirp to the signal. Before considering the dynamic chirp caused by the $C'Z$ term, we first consider a signal with a stationary sinusoidal chirp profile ${{\bar{\Omega }}_{0}}({{t}_{2}})={{\bar{\Omega }}_{c}}-{{\bar{\Omega }}_{m}}\times \cos (2\pi{t_2})$ in a cavity with $C'=0$, where ${{\bar{\Omega }}_{c}}=0.4$ and ${{\bar{\Omega }}_{m}}=0.2$ are the center and amplitude of the frequency modulation. Since ${{\bar{\Omega }}_{0}}({{t}_{2}})$ is fully within the region of $\bar{\Omega }_{0}^{2}<1$, the solution Eq.~(\ref{eq:23_chirpUbar}) is still valid. The satisfaction of stable criterion ${{\lambda }_{max}}=3\bar{\Omega }_{0}^{2}({{t}_{2}})-1<0$ depends on the value of ${{\bar{\Omega }}_{0}}({{t}_{2}})$ at each individual temporal point ${t_2}$. Here, the maximum value of $\bar{\Omega }_{0}^{2}({{t}_{2}})$ is 0.36 and larger than 1/3, the portion of the signal with $t_2\in(0.4235,0.5765)$ falls within the region of $\bar{\Omega }_{0}^{2}>{1}/{3}$ and becomes unstable. Figure \ref{fig:3_chirp_evolution}  shows the spectrograms of the signal at different $Z$. The spectrograms are generated with a moving Chebyshev gating function applied to the signal in the time domain.  At $Z=5000$, distinct sidebands have already formed at the extremum points of the frequency ${{\bar{\Omega }}_{0}}$. From $Z=5000$ to $7000$, higher order sidebands are quickly generated and widely spread out on the temporal waveform. After $Z=7000$, a new signal with a frequency lower than the original signal is generated and becomes dominant in the section as the Eckhaus instability got triggered. Eventually, the signal in the whole unstable section is replaced by thus formed new signal in the stable region and the higher order sidebands got totally suppressed in the unstable region. During this entire dynamics, the signal initially in the stable region remains unaffected.

\begin{figure}[t]
\includegraphics[width=3.4in]{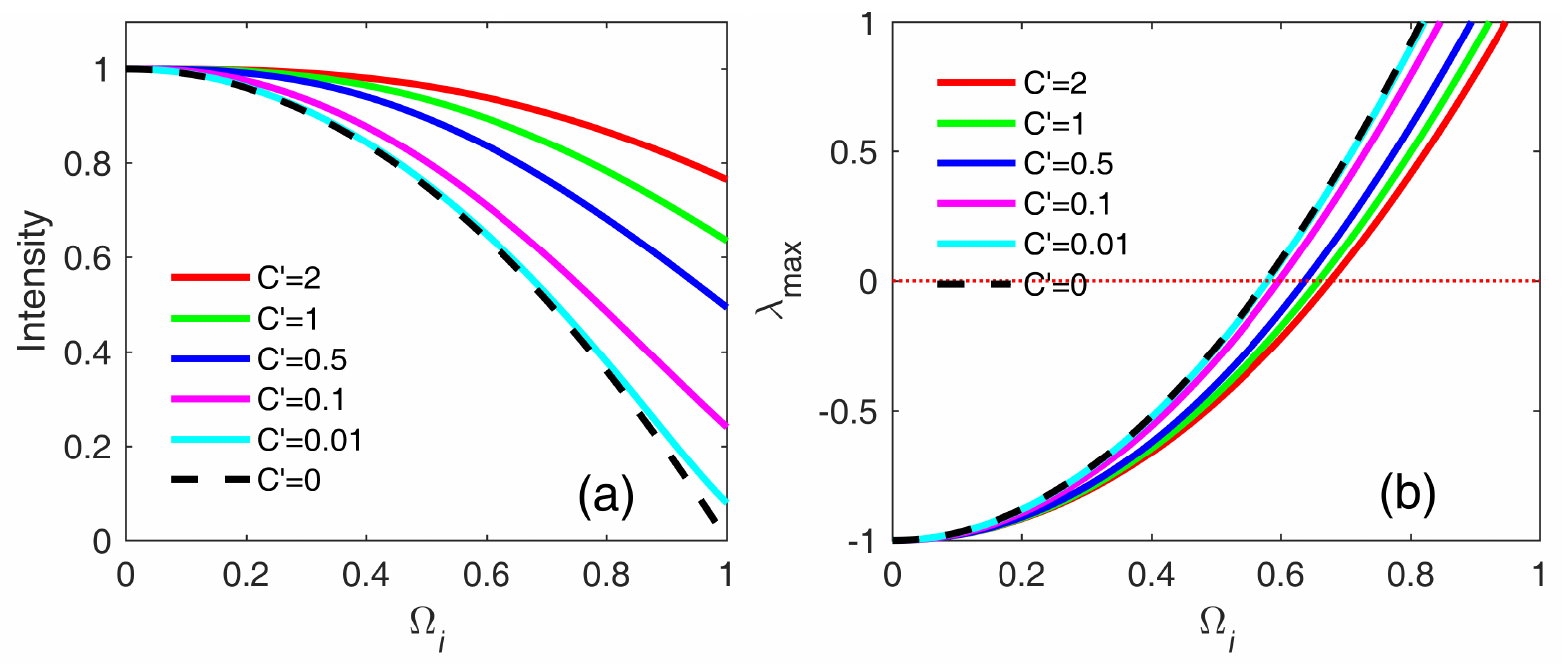}
\caption{\label{fig:4_solutionana} (a) The intensity evolution in propagation of signals with different $C'$. (b) The maximum eigenvalue $\lambda_{max}$ of the signals in (a).}
\end{figure}

From the evolution of the solution (\ref{eq:23_chirpUbar}) shown in Figs.~\ref{fig:1_sidebands}--\ref{fig:3_chirp_evolution}, it is clear that the Eckhaus instability plays a vital role in the dynamics of the signal with either a single frequency offset or a stationary frequency modulation. The portion of the signal with frequencies in the region  $\bar{\Omega }_{0}^{2}>{1}/{3}$ becomes unstable and replaced by a new signal in the stable region. But in a realistic FDML-FL, the chirp profile of the signal is not fixed but varying continuously along the propagation, which corresponds to nonzero $C'$. When $C'$ is nonzero, the system dynamics becomes more complex since the solution should be given by Eq.~(\ref{eq:12_solutionUbarorder0}). From ${{\Omega }_{i}}={C}'Z-{{\bar{\Omega }}_{0}}$, if $C'$ is nonzero, $\Omega_i$ will monotonically increase or decrease with the increase of $Z$. Such monotonic variation of $\Omega_i$ will inevitably push the signal to unstable region.
The evolutions of the intensity of the solution Eq.~(\ref{eq:12_solutionUbarorder0}) and $\lambda_{max}$ described by Eq.~(\ref{eq:22_eigenmax}) are shown in Fig.~\ref{fig:4_solutionana}. The intensity and $\lambda_{max}$ are plotted against the instantaneous frequency $\Omega_i$ since it is proportional to $Z$ when $C' \neq 0$, and identical to ${{\bar{\Omega }}_{0}}$ when $C'=0$. For the curves of $C' \neq 0$, the initial signal at $Z=0$ are assumed to have ${{\bar{\Omega }}_{0}}=0$ and $I_0=1$. The dashed curves with $C'=0$ are for the solution Eq.~(\ref{eq:23_chirpUbar}) and the corresponding stability factor. Figure \ref{fig:4_solutionana}(a) shows that for a given value of $\Omega_i$, the intensity of the signal will increase with the increasing of $C'$ especially for higher values of $\Omega_i$. When $C'<0.01$, the difference between the dynamically varying chirped solution and the stationary chirped solution indicated by the dashed curve is very small especially in the stable region. In contrast, the intensity curves of higher values of $C'$ deviates well away from the dashed curve. Besides the deviation of the intensity trace, the stability condition is also affected by the nonzero $C'$, as shown in Fig.~\ref{fig:4_solutionana}(b). When $C'$ increases from 0 to 2, the critical point of the instability has been pushed from $\Omega_i=0.577$ to $\Omega_i =0.677$. When $C'<0.01$, the critical point is almost fixed to $0.577$ which is accordant to the Eckhaus instability.

To investigate the signal dynamics and the instability in FDML-FL cavities, we consider a practical FDML-FL cavity with a length of $1$ km and a round trip time of $5$ $\mu$s. The dispersion coefficients of the fiber are $D_2=-1$ ps$^2/$km and $D_3=0.02$ ps$^3/$km. A sweeping Gaussian filter with a bandwidth of $B=0.04$ ps$^{-1}$ is driven by a sinusoidal signal ${{\omega }_{s}}(t)={{\omega }_{m}}\cos (2\pi {{f}_{0}}t)$ with repetition rate $f_0=200$ kHz and sweeping frequency range $2\omega_m=80$ ps$^{-1}$. The gain and loss coefficients of the cavity are $g_0=2.5$ km$^{-1}$ and $\sigma=0.5$ km$^{-1}$, respectively. The saturation power of the gain element is $1$ mW. Then the normalized parameters are $U=25u$, $Z=2z$, $\Omega =12.5\omega$, $t_1=0.08t$, $\epsilon =2.5\times 10^{-6}$, $t_2=\epsilon t_1$, and
\begin{eqnarray}
\label{eq:26_para_normtoreal}
&&C(t_2)=-1.6\times {10^{-4}} \times \left[\cos^2(2\pi t_2)-0.8 \cos^3(2\pi t_2) \right], \nonumber\\ 
&&C'(t_2)={10^{-4}} \times 3.2\pi \sin (4\pi t_2) \left[1-1.2\cos (2\pi t_2)\right],
\end{eqnarray}

\noindent where the maximum of $|C'|$ is 0.0019. With these normalized parameters, the dynamics of the signal can be simulated using Eq.~(\ref{eq:5_RGLEC}).

\begin{figure}[t]
\includegraphics[width=3.4in]{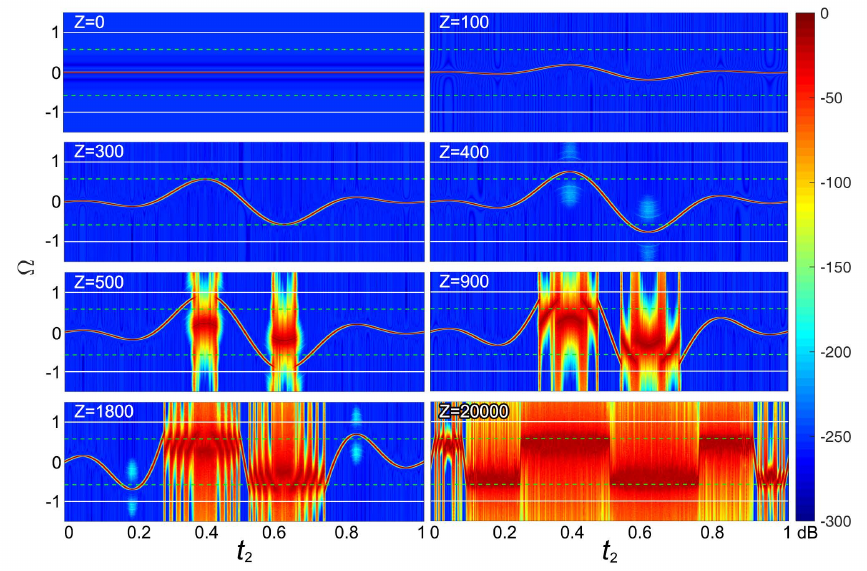}
\caption{\label{fig:5_FDML_evolution} The spectrograms of the signal in FDML-FL at $Z = 0$, 100, 300, 400, 500, 900, 1800, 20000.}
\end{figure}

We start the simulation of Eq.~(\ref{eq:5_RGLEC}) from a CW signal with ${{\bar{\Omega }}_{0}}=0$. During the propagation along $Z$, the frequency shift accumulates and continuously increases the swing range of the curve on the spectrogram, as shown in Fig.~\ref{fig:5_FDML_evolution}. In the range of $Z<300$, the spectrogram of the signal is always smooth since the entire signal is confined within the stable region of $\Omega^2<1/3$. Once the peak points of the signal cross the threshold points of $\Omega^2=1/3$ at $Z \approx 303$, sidebands start to grow in the area just outside the stable region but it is too low to be observed during early stages. At $Z=400$, the sidebands are clearly visible. As the unstable portion of the signal continuously expands, more and more higher order sidebands are also generated. At the same time, new signal generated in the stable region are as well frequency shifted towards the unstable region by the effect of dispersion. Eventually, the new signal will exit the stable region and suffer from its own Eckhaus instability. This mechanism gets repeated for the generation of new signals in the stable region, getting frequency shifted to the unstable region because of the dispersion effect and thus vanishing due to Eckhaus instability, severely distort the laser signal. After many cycles inside the cavity, e.g. at $Z=20000$, the signal almost in the entire region becomes very noisy. Most of the signal energy gets distributed near the boundary of the stable region, which means the signal experiences a very high loss when passing through the filter. The frequency of the signal is completely dispersed even for the part of the signal located at the same side of the filter. This is a great limitation to the instantaneous linewidth of FDML-FL. It should be noted that the direction of frequency shift is determined by the sign of $C'$. Thus the signals with different sign of $C'$ cluster on different sides of the filter.

In this letter, we studied the Eckhaus instability in FDML-FL. We found that the FDML-FL can be modeled by a RGLE with a frequency shifting term, which is due to the dispersion of the fiber. The fast recovering gain saturation provides a nonlinear loss to the signal. We have derived the analytical solution for the system equation and analyzed its stability. In FDML-FL, the dispersion introduces a continuous frequency shift $C'$ to the signal, which will unavoidably push the frequency outside the stable region of $\Omega^2<1/3$. If $C'$ is large, the stable region on frequency domain will be slightly enlarged. By considering practical parameter values, we numerically showed the repeatedly triggering of Eckhaus instability in FDML-FL cavities by the endless frequency shifting. Such mechanism is the root cause for the high frequency fluctuations of the signal that limits the signal quality of the FDML-FL.

We acknowledge the support of Research Grant Council of Hong Kong SAR (PolyU5263/13E, PolyU152144/15E), The Hong Kong Polytechnic University (1-ZVGB), National Science Foundation of China (NSFC) (61475131) and Shenzhen Science and Technology Innovation Commission (JCYJ20160331141313917).

\bibliography{PRL_Eckhaus}

\end{document}